\documentclass[a4paper]{jpconf}

\usepackage{graphicx}
\usepackage[usenames]{color}
\usepackage{subeqn}
\usepackage{ulem} 
\usepackage{textcomp}

\begin{document}

\newcommand{\cred}[1]{{\color{BrickRed}#1}}
\newcommand{\kpar}{k_\parallel}
\newcommand{\lint}[3]{\int\limits_{#1}^{#2}d{#3}\,}

\title{Maximizing phonon thermal conductance for ballistic membranes}

\author{T. K\"uhn and I. J. Maasilta}

\address{Department of Physics, Nanoscience Center, P. O. Box 35, FIN-40014, University of Jyv\"askyl\"a, Finland }

\ead{kuehn@jyu.fi, maasilta@phys.jyu.fi}

\begin{abstract}

At low temperatures, phonon scattering can become so weak that phonon transport becomes ballistic. 
We calculate the ballistic phonon conductance $G$ for membranes using elasticity theory, considering 
the transition from three to two dimensions. We discuss the temperature and thickness dependence 
and especially concentrate on the issue of material parameters. For all membrane 
thicknesses, the best conductors have, counter-intuitively, the {\em lowest} speed of sound.  
\end{abstract}

\section{Introduction}

Recent advances in nanoscience are expanding the limits of phononic thermal transport, both in 
the low conductance and high conductance side \cite{kim}. The highest thermal conductance of a 
particular material is achieved, when the phonons do not scatter at all; this is the ballistic 
transport limit. At room temperature the ballistic limit is quite hard to achieve, but at low 
temperatures (a few K and below) phonon-phonon and phonon-impurity scattering become so weak 
that ballistic transport can be observed \cite{vongut,klitsner}, and the emitted  power has the 
typical Stefan-Boltzmann form \cite{wolfe}
\begin{equation}
P=A\sigma T^4\,, \hspace{5mm}\mbox{where for phonons}\hspace{5mm}
\sigma=\pi^5k_B^4/(15h^3)\Sigma e_i/c_i^2\,,\label{eqn_stefan_boltzmann_law}
\end{equation}
summing over the different phonon modes with speeds 
of sound $c_i$ and radiator emissivities $e_i$.
   
In addition to bulk samples,  ballistic phonon transport was also observed for thin
amorphous SiN$_\mathrm{x}$ membranes  recently \cite{sron}. 
At low temperatures, the dominant thermal phonon wavelength becomes comparable to
the membrane thickness (for a 1{\textmu}m $\mathrm{SiN}_\mathrm{x}$ membrane this happens at
$\approx$100mK) leading to an effectively 2D phonon gas.
We have recently considered ballistic and diffusive
thermal conduction in thin membranes \cite{kmnima,prepr,TLSprepr}, 
spanning the transition from the 3D to the fully 2D limit, 
using the correct low-frequency modes known from elasticity theory (Lamb modes). Here we explicitly 
discuss the dependence of ballistic thermal conduction on materials parameters. The simple question we 
would like to answer is: what kind of material is the best (worst) ballistic thermal conductor 
for any thickness of the membrane, and can we even define a material-only dependent thermal 
conductivity that does not depend on the sample dimensions?  
       
For the more typical cases where phonon transport is not ballistic, but limited by different 
scattering mechanisms in the sample, it is possible to define a material dependent local thermal 
conductivity $\kappa$, which depends on the specific heat $c_V$, average
speed of sound $c$ and phonon 
mean free path $\Lambda$ as $\kappa=1/3c_Vc\Lambda$. For 3D crystals, $c_V$ is proportional to 
$1/c^3$, but because $\Lambda$ scales as $\Lambda \propto c^n$, where $n=3,4,5$ for 
phonon-electron, impurity, and phonon-phonon scattering, respectively, $\kappa$ always grows 
with higher speed of sound \cite{wolfe}.
This is the intuitive result that high speed of sound materials are 
better thermal conductors (like diamond). However, for the case of ballistic conduction this 
is not true anymore. The thermal conductance $G$ 
(one cannot define a local $\kappa$ anymore) is calculated from the total
power $P$ that is radiated from an object as $G=dP/dT$. $G$ actually 
has a dependence $G \propto 1/c^2$, as can be 
seen from the Stefan-Boltzmann law, Eq.\ (\ref{eqn_stefan_boltzmann_law}).
The material with the {\em lowest} speed of sound is 
the best thermal conductor! In the following we will
investigate what is the situation for thin membranes. 

\section{Membrane eigenmodes}

In isotropic 3D bulk systems there are 
three independent phonon modes, two transversally
and one longitudinally polarized, with sound velocities $c_t$ and $c_l$, respectively. 
In the presence of boundaries, the bulk phonon modes
couple to each other and form a new set of eigenmodes, which in the case of a free standing
membrane are horizontal shear modes ($h$) and symmetric ($s$) and antisymmetric
($a$) Lamb modes \cite{Auld}.
The frequency $\omega_h$ for the $h$ modes is simply 
$\omega_h= c_t\sqrt{\kpar^2+(m\pi/d)^2}$,
where $\kpar$ is the wave vector component parallel to membrane surfaces,
$d$ is the membrane thickness and the integer $m$ is the branch number. The dispersion
relations of the $s$ and $a$ modes 
cannot be given in a closed form, but have to be 
calculated numerically \cite{PhysRevB}. The lowest three branches, dominant for thin membranes at 
low temperatures, have the low frequency expressions
$\omega_{h,0} = c_t\kpar$, $\omega_{s,0} = c_s\kpar$ and 
$\omega_{a,0} = \frac{\hbar}{2m^\star}\kpar^2$,
%
%
%
%
%
with the effective sound velocity $c_s=2c_t\sqrt{(c_l^2-c_t^2)/c_l^2}$ of the $s$ mode 
and the effective mass $m^\star=\hbar\left[2c_td\sqrt{(c_l^2-c_t^2)/3c_l^2}\right]^{-1}$
of the $a$ mode ``particle'' \cite{PhysRevB}.
This lowest 
$a$ mode 
has a quadratic 
dispersion instead of the usual phonon-like linear one, 
and is mostly 
responsible for the non-trivial behavior of thin membranes at low temperatures. 

\section{Ballistic conduction}

\begin{figure}[h]
\begin{minipage}[t][68mm][t]{18pc}
\includegraphics[width=18pc]{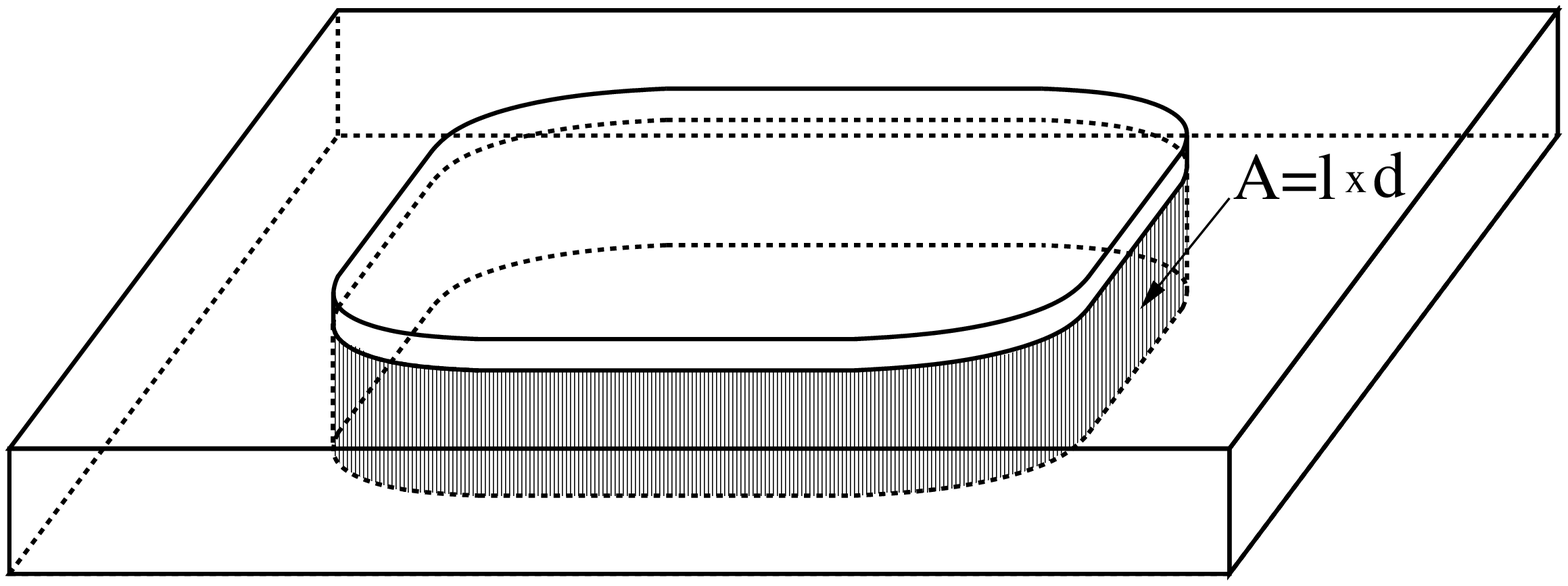}
\caption{\label{fig_fig1} Schematic of the considered geometry.
A thin metal film heats
the part of the membrane directly below.
The radiating area is $A=l\times d$}
\end{minipage}\hspace{2pc}
\begin{minipage}[t][68mm][t]{18pc}
\includegraphics[width=23pc]{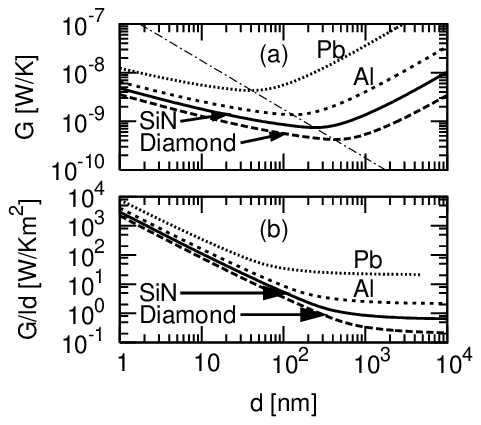}
\caption{\label{fig_fig2}
(a) Ballistic conductance $G$ and
(b) Ballistic conductance per unit area $G/ld$ as function of membrane
thickness $d$. The dash-dotted line in (a) is $\propto 1/d$.}
\end{minipage} 
\end{figure}

Figure \ref{fig_fig1} shows a schematic of the system in consideration. A thin metal film
of perimeter $l$ heats the membrane directly below \cite{prepr}.
To simplify the discussion, we assume
that the hot phonons from the heater have a thermal distribution and are radiated from its 
perimeter into the membrane, similar to black body radiation. We further assume that
no radiation is backscattered. 
With these assumptions the total heat flow out of the detector is
\begin{equation}
P=\frac{l}{2\pi^2}\sum_{\sigma,m}\lint{0}{\infty}{k_\parallel} k_\parallel
     \hbar\omega_{\sigma,m}\left|\frac{\partial\omega_{\sigma,m}}{\partial k_\parallel}
     \right|n(\omega,T),\label{eqn_P_of_T}
\end{equation}
where $n(\omega,T)$ is the Bose-Einstein distribution 
and $\sigma$ and $m$ are the mode and branch indices \cite{note2}. If enough branches
are used, the 3D to 2D transition can be computed from this expression.
If the membrane is 
thin and temperature low ($Td\ll\hbar c_t/2k_B$), only the 
lowest branch of each mode is occupied ($m$=0),
and we are fully in the 2D limit, in which case
we get from Eq.\ (\ref{eqn_P_of_T})
%
\begin{equation}
P_\mathrm{2D} 
 =  \frac{l\hbar}{2\pi^2}\left[\!\left(\frac{1}{c_t}+\frac{1}{c_s}\right)\!
      \Gamma(3)\zeta(3)\left(\frac{k_BT}{\hbar}\right)^3\right.\nonumber\\
     \left.+\sqrt{\frac{2m^*}{\hbar}}\Gamma\!\left(\frac{5}{2}\right)\!
      \zeta\!\left(\frac{5}{2}\right)
      \!\left(\frac{k_BT}{\hbar}\right)^{\!5/2}\right]\,.\label{eqn_P_2D}
\end{equation}
Note that the effective mass of the lowest $a$ mode depends on the membrane thickness and hence
in the 2D limit $P\propto 1/\sqrt{d}$. In the 3D limit $Td\gg\hbar c_t/2k_B$,
the dominant phonon wavelength is much smaller than $d$, 
leading to decoupling of the longitudinal and transversal modes and
\begin{eqnarray}
P_\mathrm{3D}
 &=& \frac{\pi^2ld\hbar}{120}\!\left(\frac{2}{c_t^2}+\frac{1}{c_l^2}\right)
     \!\left(\frac{k_BT}{\hbar}\right)^4\,.\label{eqn_P_3D}
\end{eqnarray}
As expected, $P_\mathrm{3D}\propto d$. 
This means that at a fixed temperature the radiated power will first decrease with decreasing $d$, 
then reach a global minimum and will increase again, if we decrease $d$ further.
The minimum is approximately at the
2D-3D crossover thickness $d_C\equiv \hbar c_t/(2k_B T)$.

\section{Numerical results}

Using the 300 lowest modes, we have numerically computed the ballistic phonon conductance $G$ for a 
range of parameters that spans the transition from 2D to 3D. Fig.\ \ref{fig_fig2} (a)
shows the computed $G$ for 
$T$=100 mK and $l$=1.6 {\textmu}m
as a function of the membrane thickness $d$ for four different materials: 
Pb ($c_t=1025$ m/s, $c_l=2370$ m/s), Al ($c_t=3280$ m/s, $c_l=6580$ m/s), 
SiN ($c_t=6200$ m/s, $c_l=10 300$ m/s), and poly-diamond ($c_t=10 900$ m/s, $c_l=18 100$ m/s). 
The values of $c$ are calculated values for the isotropic case, based on measured single crystal 
elastic constants \cite{simmons}. All materials have a physically relevant critical thickness, 
ranging from $d_C= 40$ nm for Pb to $d_C=400$ nm in diamond, below which $G$ increases with 
decreasing thickness. The material with highest $G$ is Pb, as it has the lowest speeds of sound, 
and diamond is the poorest conductor. In the 2D limit, where $G\propto 1/\sqrt{c}$, the differences 
between the materials are not as strong as in the 3D limit. One could say that all materials 
start to resemble each other in the ultrathin range $d < 10$ nm.    

On the other hand, in the 3D limit, 
$G$ increases linearly with $d$,
the common sense result. 
Therefore, we could try to define a materials parameter $G/ld$, the conductance per unit emission 
area, for all thicknesses. This is plotted in Fig.\ \ref{fig_fig2} (b).
As we see, in the 3D limit $G/ld$ is not 
a function of $d$, and is truly a materials parameter. However, when the membrane becomes thinner, 
$G/ld$ starts increasing strongly. This means that thin membranes are actually more effective thermal 
conductors, per unit emission area. This increase in ballistic "conductivity" is striking: 
for diamond it can be four orders of magnitude from $d= 1 \mu$m to $d=1$ nm. 
Therefore, a way of increasing the ballistic thermal conductance in a given volume would
be to use a structure composed of many parallel thin membranes of a material with low speed
of sound.

\begin{figure}[h]
\begin{minipage}[t][78mm][t]{18pc}
\includegraphics[width=18pc]{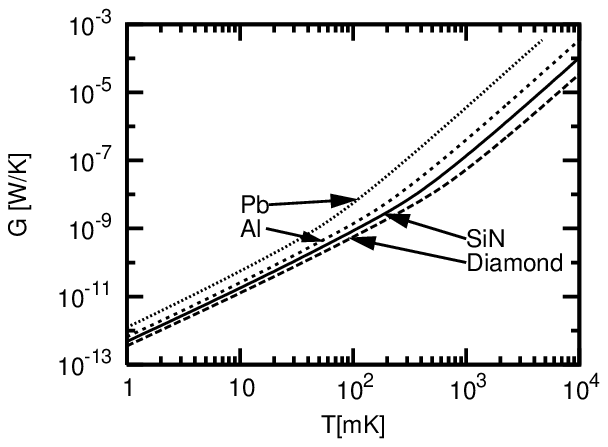}
\caption{\label{fig_fig3}Ballistic conductance $G$ as function of temperature. In the
3D limit, $G\propto T^3$, while in the 2D limit we have $G\propto T^{3/2}$.}
\end{minipage}\hspace{2pc}%
\begin{minipage}[t][78mm][t]{18pc}
\includegraphics[width=18pc]{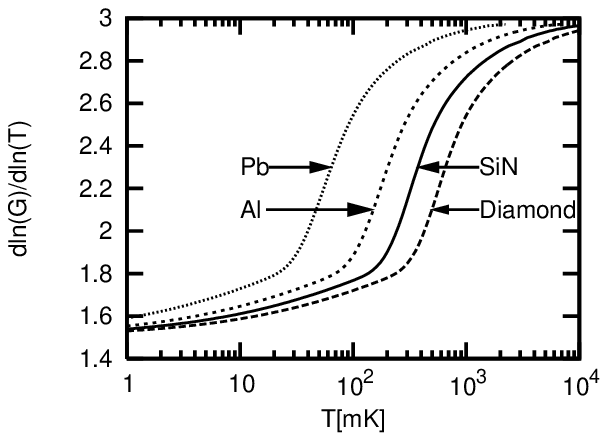}
\caption{\label{fig_fig4}Logarithmic derivative of the ballistic conductance 
$\alpha\equiv d\ln(G)/d\ln(T)$ as function of temperature. In the 3D limit $\alpha=3$,
while in the 2D limit $\alpha \to 3/2$.}
\end{minipage} 
\end{figure}

The dependence of $G$ on temperature is plotted in Fig.\ \ref{fig_fig3} for $d=100$ nm. In 
the 3D limit, 
$G$ is naturally proportional to $T^3$. In the 2D limit however, the power of the temperature
dependence does not just drop by one, but instead we have $G \propto T^{3/2}$,
and the crossover temperature is between 50 .. 400 mK for $d=100$ nm. To give a better
picture of how the power of the temperature dependence of $G$ changes with temperature,
we plot the logarithmic derivative $d\ln(G)/d\ln(T)$ of $G$ in Fig.\ \ref{fig_fig4}.

\section{Conclusions}

For a given temperature, ballistic thermal conduction is maximized by choosing a material 
with low speed of sound, and by using several parallel thin membranes, instead of a bulk 
slab of the same total thickness. A bulk slab of diamond of thickness 30 $\mu$m has the 
same conductance as a 2 nm thick lead membrane. The lead membrane is thus about 15 000 more 
effective conductor. Applications in cooling ultrasensitive thermal detectors can be envisioned.

\section*{Acknowledgments}

This work was supported by the Academy of Finland project Nos. 118665 and 118231.

\section*{References}


\begin{thebibliography}{10}
\bibitem{note2} Note that the prefactor is $\frac{l}{2\pi^2}$ instead of the factor $\frac{l}{8\pi}$
that was mistakenly given in Ref.\ \cite{kmnima}.
\end{thebibliography}
\end{document}